\documentclass[twocolumn,prb,aps]{revtex4}

\begin{document}


\title{Plasma-Like Negative Capacitance in Nano-Colloids}

\author{J.~Shulman}
\author{S.~Tsui}
\author{F.~Chen}
\author{Y.~Y.~Xue}
\affiliation{Department of Physics and Texas Center for Superconductivity,
University of Houston, 202 Houston Science Center, Houston, Texas 77204-5002}
\author{C.~W.~Chu}
\affiliation{Department of Physics and Texas Center for Superconductivity,
University of Houston, 202 Houston Science Center, Houston, Texas 77204-5002;
Lawrence Berkeley National Laboratory, 1 Cyclotron Road,
Berkeley, California 94720; and Hong Kong University of Science and Technology,
Hong Kong}

\date{\today}

\begin{abstract}
A negative capacitance has been observed in a nano-colloid between
0.1 and $10^{-5}$ Hz. The response is linear over a broad range of
conditions. The low-$\omega$ dispersions of both the resistance and
capacitance are consistent with the free-carrier plasma model, while
the transient behavior demonstrates an unusual energy storage
mechanism. A collective excitation, therefore, is suggested.
\end{abstract}

\maketitle

The origin of the negative capacitance (NC) observed in many systems with mesostructures has
been debated since 1988.\cite{wer,lem} Experimentally, this reactive response is probed by
the phase shift between the applied $ac$ voltage, $V_{ac}$, and the resulting current,
$I_{ac}$, at a frequency $\omega$. A negative phase-angle between $I_{ac}$ and $V_{ac}$,
therefore, represents a NC. Jonscher,\cite{jon} Ershov {\it et al.},\cite{ers} and others,
however,
repeatedly emphasize that the clue to these unusual observations lies in the equivalent
negative transient current $\Delta I(t > 0) = [I(t)-I(\infty)]$ after a positive voltage jump
$\Delta V = V(0^{+})-V(0^{-})$ at $t = 0$. Such a retarded current response, consequentially,
is attributed to the inertia of the associated process. Penin\cite{pen} refers to such a
process, which is characterized by the equation $dI/dt+I/\tau = bV$ with parameters $\tau$
and $b$, as a $\tau$ system. The equation, it should be pointed out, is actually that of a
free-carrier plasma with the solution of $I_{ac}/V_{ac} = 1/R-i\omega C = b/(i\omega +1/\tau)
= b \cdot (1/\tau -i\omega)/(\omega^{2}+1/\tau^{2})$, where $R$ is the resistance, $C$ is the
capacitance, and $1/\tau$ represents the plasma damping. It is therefore interesting to note
that the proper mesostructures may theoretically lead to a negative dielectric
response,\cite{pen2} and any negative static $\epsilon '$ (which leads to NC at $\omega = 0)$
can be described as ghost-plasmons.\cite{tak}

A true plasma-like response, however, should also have a linear
$I-V$ characteristic as well as follow the causality correlation
between $R$ and $C$. The observed $\Delta I(0^{+})/\Delta V$, for
example, should include a component of $\Delta I_1 = 1/R_{\omega =
\infty} - 1/R_{\omega = 0} = -b\tau$ from the $R$-dispersion in
addition to the term of $\Delta I_2 = 2/\pi
\int\limits_{0}^{\infty}[C(\omega) - C(\infty)]d\omega = -b\tau$,
which was derived in Ref. \onlinecite{ers} from the
$\omega$-dependent $C$ alone. The high nonlinearity as well as
limited data resolution in previously published data, unfortunately,
makes a full identification difficult.\cite{wer,lem,jon,ers,pen} In
some cases (such as the impact ionization of gas in Ref.
\onlinecite{pen}), such a plasma interpretation is obviously an
oversimplified approximation. Here we report the NC below 1 Hz in a
giant electrorheological (ER) fluid.\cite{che} The $I-V$
characteristic is linear across a broad $V_{ac}$ range, and all
features, {\it e.g.} $C(\omega)$, $\Delta I$, and the $R(\omega)$
contribution to it, are consistent with a plasma-like excitation
with an extremely long $\tau$.

The ER fluid is a colloid of silicone oil and 20 nm urea-coated
Ba$_{0.8}$Rb$_{0.4}$TiO(C$_{2}$O$_{4}$)$_{2}$ nanoparticles.\cite{che,wen} The nanoparticle to
oil ratio is 10 g: 3 ml. The capacitor cells were constructed of two parallel gold or copper
electrodes with dimensions of 6 mm $\times$ 13 mm and a gap distance of 0.1 mm. The $ac$
measurements were performed via a serial connected $ac$ voltage source, capacitor cell, and
current meter.  Only $dc$ couplings were present. The phase angle was verified with
resistors and capacitors of known values. Both the systematic error and the resolution of
the phase angle are within $0.05^{\circ}$. Details of both the ER fluid preparation\cite{wen}
and the measurements\cite{che} have been reported before. The same hardware was used in the
transient measurements with the pulse output and the voltage reading being synchronized
within $10^{-4}$ sec.

The effective $R(f)$ and $C(f)$, where $f = \omega/2\pi$, with an
$ac$ excitation signal $V_{ac}$ of 1 V between 0.0005 and 100 Hz is
shown in Fig.~\ref{fig:fig1}a and b under various $dc$ biases
$V_{dc}$. With zero bias, the capacitance increases with the
decrease of $\omega$. This behavior is typical of a disordered
system and is referred to as the universal dielectric response
(UDR).\cite{dyr} With the application of a $dc$ bias, the low
frequency capacitance becomes negative while the UDR characteristics
are retained at higher frequencies. The capacitance changes from
positive to negative around 0.1 Hz and is independent of the bias
for $V_{dc} \ge 300$ V. This excellent linearity is verified by the
$V_{ac}$-independence of $C$ up to 100 V peak-to-peak (inset,
Fig.~\ref{fig:fig1}a). It is interesting to note that the observed
dispersion can be roughly separated into two parts: a relatively
small UDR-like $C(\omega) > 0$ above 1 Hz and a plasma-like NC
(solid line in Fig.~\ref{fig:fig1}a), which dominates the off-phase
part below 0.01 Hz. Presently, the exact role of the $dc$ bias in
the induced NC is unclear.  It is believed that excess carriers are
injected into the system with the application of bias, resulting in
the plasma-like characteristics.

The transient current should be negative around and after 0.1 sec. based on the measured
$C(\omega)$ and linearity. We demonstrated this by directly measuring $\Delta I(t)$ for a
voltage sequence of 340 V/350 V/360 V/350 V with a uniform time interval of 1032 sec. The
trend can even be observed in the raw data, although with a significant baseline shift
(inset, Fig.~\ref{fig:fig2}). To obtain a quantitative result, a smooth fit of
$I_{base} = \sqrt{m + n \cdot t^{2}}$ (the thick line in the inset) was used as the baseline,
where $m$ and $n$ are fitting parameters. The transient currents at 350 V during a 10 hr
period (about 7 cycles) are then realigned by subtracting the start times of each cycle
(grey traces in Fig.~\ref{fig:fig2}), and the average is plotted as the black line with its
width representing the standard error. The $\Delta I$ is indeed negative after 1--2 ms, and
the baseline shifts do not affect the conclusion (Fig.~\ref{fig:fig2}).

For further comparison with the plasma model, both the nonlinearity
and the $\omega$-dependence in the $R$-section were carefully
reviewed.  The relatively small $\Delta I(t)$, $< 1$\% of the
corresponding $I(\infty)$, requires better resolution than the data
in Fig.~\ref{fig:fig1}b, where the long-time shift limits the data
accuracy (the data acquisition requires a few-week period). Local
$ac$ $I(V)$ loops with the same $V_{dc}$ of 350 V and a peak-to-peak
$|V_{ac}|$ of 20 V, therefore, were measured at 0.3 and 0.001 Hz,
respectively (Fig.~\ref{fig:fig3}). It should be pointed out that
the contribution from both the $\omega > 0.3$ Hz and the $\omega <
0.001$ Hz parts should be negligible within the time window of
1--1000 sec. The relevant contributions from the passive $R$-channel
at $t = 0^+$ and $\infty$, therefore, may be deduced from $R$(0.3
Hz) and $R$(0.001 Hz), respectively. The passive and reactive
components, {\it i.e.} the $R$ and the $C$ of the sample, are
represented by the non-hysteretic and the hysteretic parts of the
$I-V$ loops, respectively (lower inset, Fig.~\ref{fig:fig3}). The
$R$ component, therefore, is taken as the average of the $V$
increase- and $V$ decrease-branches (Fig.~\ref{fig:fig3}). Two
features immediately emerge: a) the $I-V$ is rather linear within
the experimental resolution of a few tenths of a percent over a 20 V
range; and b) there is a significant $\omega$-dependence of $R$,
which contributes to $\Delta I$. The observed $\Delta V \cdot
(1/R_{0.3~\rm Hz}-1/R_{1~\rm mHz}) \approx 0.08$ $\mu$A
(Fig.~\ref{fig:fig3}) is roughly half of the transient current
directly measured (Fig.~\ref{fig:fig2}). It is interesting to note
that this is exactly what is expected from the plasma model: $\Delta
I_1 = -b\tau = \Delta I_2 = \Delta I/2$. Despite the large
background, the plasma-like $R$-contribution is demonstrated in the
observed NC.

Another interesting issue is the energy balance in such a system.
The negative transient current $\Delta I$ actually flows against the
field for a positive $\Delta V$, corresponding to an energy
storage/conversion in the linear system (Fig.~\ref{fig:fig3}).
Unlike ordinary capacitors, energy is released from the material in
the $V$-increase branch, but stored during the $V$-decrease branch.
This is, again, a characteristic of a plasma-like excitation. The
extremely long phase coherence period, $\tau \sim 141$ s, however,
suggests that the excitation may be soliton-like in nature.  An
unusual surface plasma and quantum capacitance\cite{dul} are
possible candidates.

In summary, we have observed a negative capacitance in an ER fluid.
Linearity is observed over broad ranges of $V_{dc}$ and $V_{ac}$.
The dispersions in both $C$- and $R$-channels and the associated
energy storage/conversion demonstrate that plasma-like excitations
are present. This demonstrates that dielectric properties can be
drastically changed through mesostructures, and opens the door for
novel nano-materials.

\begin{acknowledgments}
The authors thank Prof. W. Wen for the ER fluid samples. The work in Houston is supported
in part by
AFOSR Award No. FA9550-05-1-0447, the T.~L.~L. Temple Foundation, the John J. and Rebecca
Moores Endowment, the Strategic Partnership for Research in Nanotechnology through AFOSR
Award No. FA9550-06-1-0401, and the State of Texas
through the Texas Center for
Superconductivity at the University of Houston; and at Lawrence Berkeley
Laboratory by the Director, Office of Science, Office of Basic Energy Sciences,
Division of Materials Sciences and Engineering of the U.S. Department of Energy
under Contract No. DE-AC03-76SF00098.
\end{acknowledgments}



\begin{figure}

\caption{\label{fig:fig1}a) $C(f)$ at $V_{dc}$ of 0 (circles), 3.5
(squares) and 5 kV/mm (triangles). The line is the fit for free
carrier plasma with $b = 7.2 \cdot 10^{-11}$  $s^{-2}$ and $\tau =
141$ s; Inset: the deduced $C$ vs. $V_{ac}$ at $V_{dc} = 350$ V and
10 mHz. b) Differential $R$ at $V_{dc}$ of 500 (squares) and 350 V
(triangles).}

\caption{\label{fig:fig2}$\Delta I$ during a voltage serial of 340 V/350 V/360 V/350 V \dots
~Inset: the raw data of $I$. Thick solid line: the assumed base line. The positive part
associated with the geometric capacitance is absent due to the time window used.}

\caption{\label{fig:fig3}The non-hysteretic parts of the $ac$ loop at $V_{dc} = 350$ V and
$V_{ac} = 10$ V. Solid line: at 1 mHz; dashed line: 0.3 Hz. Bottom inset: the raw I-V loops
at 1 mHz. Top inset: The power output after a positive 10 V jump.}

\end{figure}


\begin{thebibliography}{99}

\bibitem{wer} J. Werner, A. F. J. Levi, R. T. Tung, M. Anzlowar, and M. Pinto, Phys. Rev.
Lett. {\bf 60}, 53 (1988).

\bibitem{lem} For example, F. Lemmi and N. M. Johnson, Appl. Phys. Lett. {\bf 74}, 251 (1999);
H. C. F. Martens, J. N. Huiberts, and P. W. M. Blom, {\it ibid.} {\bf 77}, 1852 (2000);
L. Bakueva, G. Konstantatos, S. Musikhin, H. E. Ruda and A. Shik, {\it ibid.} {\bf 85}, 3567
(2004); L. S. C. Pingree, B. J. Scott, M. T. Russell, T. J. Marks, and M. C. Hersam,
{\it ibid.} {\bf 86}, 73509 (2005).

\bibitem{jon} A. K. Jonscher, J. Chem. Soc. Faraday Trans. II {\bf 82}, 75 (1986).

\bibitem{ers} M. Ershov, H. C. Liu, L. Li, M. Buchanan, Z. R. Wasilewski, and A. K. Jonscher,
IEEE Trans. Electron Devices {\bf 45}, 2196 (1998).

\bibitem{pen} N. A. Penin, Semiconductors {\bf 30}, 340 (1996).

\bibitem{pen2} J. B. Pendry, L. Mart\'in-Moreno, and F. J. Garcia-Vidal, Science {\bf 305},
847 (2004).

\bibitem{tak} K. Takayanagi and E. Lipparini, Phys. Rev. B {\bf 56}, 4872 (1997).

\bibitem{che} F. Chen, J. Shulman, S. Tsui, Y. Y. Xue, W. Wen, P. Sheng, and C. W. Chu,
Phil. Mag. {\bf 86}, 2393 (2006).

\bibitem{wen} W. Wen, X. Huang, S. Yang, K. Lu, and P. Sheng, Nature Materials
{\bf 2}, 727 (2003).

\bibitem{dyr} J. C. Dyre and T. B. Schr{\o}der, Rev. Mod. Phys. {\bf 72}, 873 (2000).

\bibitem{dul} S. C. Dultz and H. W. Jiang, Phys. Rev. Lett. {\bf 84}, 4689 (2000).

\end{thebibliography}

\end{document}